\documentclass[preprint,prc,showpacs,preprintnumbers,amsmath,amssymb,floatfix]{revtex4}
\usepackage{amsmath}
\usepackage{graphicx,latexsym,amssymb, epsfig}
\usepackage{multirow,amsmath,array,booktabs}
\usepackage{subfigure}
\usepackage{color}
\usepackage{bm}
\usepackage[figuresright]{rotating}
\usepackage[section]{placeins}
\usepackage{slashed}
\usepackage{graphicx}
\usepackage{tikz}

\newcommand{\nc}{\newcommand}       
\nc{\vc}[1] {\mbox{\boldmath $#1$}} 
\nc{\del}       {\partial}              
\nc{\bra}       {\langle}               
\nc{\ket}       {\rangle}               
\nc{\bras}[1]   {\langle #1|}           
\nc{\kets}[1]   {|#1\rangle}            
\nc{\mapleft}[1]{           
 \smash{\mathop{\,          %
  \hbox to 1.5cm{\rightarrowfill}\, }\limits_{#1}}}
\nc{\beq}     {\begin{eqnarray}} \nc{\eeq}    {\end{eqnarray}}
\nc{\nn}      {\\\nonumber} \nc{\vs}      {\vspace{-0.275cm}}
\nc{\fra}    {\frac{1}{2}}
\nc{\mb}        {\mathbf}

\begin{document}

\title{Neutron drop trapped in axially deformed external fields}

\author{Xiaohan Ding}
\affiliation{School of Physics, Nankai University, Tianjin 300071,  China}
\author{Jinniu Hu}
\email{hujinniu@nankai.edu.cn}
\affiliation{School of Physics, Nankai University, Tianjin 300071,  China}
\author{Ying Zhang}
\email{yzhangjcnp@tju.edu.cn}
\affiliation{Department of Physics, Faculty of Science, Tianjin University, Tianjin 300072, China}
\author{Hong Shen}
\email{songtc@nankai.edu.cn}
\affiliation{School of Physics, Nankai University, Tianjin 300071,  China}


\date{\today}
\begin{abstract}
The neutron drop is firstly investigated in an axially symmetric harmonic oscillator (ASHO) field, whose potential strengths of different directions can be controlled artificially. The shape of the neutron drop will change from spherical to oblate or prolate according to  the anisotropy of the external field. 
With the potential strength increasing in the axial direction, the neutron prefers to occupy the orbital perpendicular to the symmetry axis. On the contrary, the neutron likes to stay in the orbital parallel to the symmetry axis when the potential strength increases in the radial direction.  Meanwhile, when the potential strength of one direction disappears, the neutron drop cannot bind together.  These investigations are not only helpful to simulate the properties of neutrons in finite nuclei but also provide the theoretical predictions to the future artificial operations on the nuclei like the ultracold atom system, for a deeper realization of quantum many-body systems.    
\end{abstract}



\maketitle

\section{Introduction}
The interacting Fermi or Bose gas trapped in an external field has been widely investigated for the ultracold atoms~\cite{anderson95,bradley95,davis95, demarco99,schreck01,truscott01}. Their interactions could be tuned through the Feshbach resonances, which provided good opportunities to investigate the correlations in quantum many-body systems. Nowadays, the external field produced by the optical lattice is usually simplified as a harmonic oscillator (HO) potential~\cite{bloch08,giorgini08,bloch09,blume12}.  When the potential strengths have a very big difference in different directions, the system of ultracold atoms could even have the quasi-one- or quasi-two-dimensional shape~\cite{butts97,odelin99,idziaszek05}. 

Similar to the atoms, the nucleus is also a complex quantum many-body systems, except that it is self-bound by the strong interaction among nucleons~\cite{zinner13}.  A nucleus can be simply approximated as many non-interacting nucleons confined in an external field, such as HO or Wood-Saxon potential~\cite{ring80}. The shell evaluations of deformed nuclei were reasonably explained by the Nilsson model with an axially symmetric HO (ASHO) potential~\cite{nilsson55}. Properties of superdeformed and hyperdeformed nuclei can also be explained by this potential~\cite{nazarewicz92}, even for the $\alpha$ linear chain configuration~\cite{freer95}. 

Recently, the neutron drop, as an artificial nuclear system, is becoming a hot research object, where the interacting neutrons are trapped in external fields~\cite{pudliner96,gandolfi11}. This system is helpful to simulate the properties of the neutron-rich system far from the $\beta$-stability line and the inhomogeneous structure of the neutron star crust.  Furthermore, the results of neutron drop given by the \textit{ab initio} methods~\cite{pudliner96,gandolfi11,potter14,shen18} could be used to calibrate the effective Hamiltonians in the nuclear density functional theory (DFT)~\cite{zhao16,shen19}.
So far, the ground state properties, pairing correlation, tensor effect, shell evaluation of neutron drops have been widely investigated in the framework of DFT and \textit{ab initio} methods~\cite{pudliner96,gandolfi11,potter14,bonnard18,shen18a,ge2020a,ge2020b}. In these investigations, all the neutron drops were trapped in spherical external fields, mostly in the isotropic HO potentials. Nuclei could have manifold shapes and structures, such as, spherical, prolate, oblate, and so on due to the change of the single-particle levels.  In some nucleus, the central density may be less than the outer region,  which forms the `bubble' structure~\cite{shukla14}.  In practice, the strong magnetic field may exist in the neutron star and heavy-ion collision process, which also influences the deformation of nuclear many-body systems~\cite{oertel17}. 

In this work, we will investigate the neutron drops trapped in deformed external fields, whose potential strengths of different directions can be controlled artificially,  and show their influences on the properties of neutron drops. These results will be compared to the non-interacting neutrons confined in the same potential to show the effect of nucleon-nucleon (NN) interaction.  Similar to the ultracold atom system, the neutron drop will be trapped in an ASHO potential.  

\section{The neutron drop in the axially deformed Skyrme Hartree-Fock-Bogoliubov model }
Apart from the deformation effect, the pairing correlation was found to have a non-negligible contribution to the neutron drop in DFT~\cite{ge2020a}. Therefore,  in this work, the neutron drop will be studied by using the axially deformed Skyrme Hartree-Fock-Bogoliubov (HFB) approach, including both the pairing and deformation effects~\cite{bender03}.
{With a zero-range Skyrme interaction in the particle-hole channel and a delta-pairing force, the typical HFB equation can be given as:
\begin{equation}\label{HFBeqn}
	\sum\limits_{\sigma '} {\left( {\begin{array}{*{20}{c}}
				{h\left( {\vec r,\sigma ,\sigma '} \right)}&{\tilde h\left( {\vec r,\sigma ,\sigma '} \right)}\\
				{\tilde h\left( {\vec r,\sigma ,\sigma '} \right)}&{ - h\left( {\vec r,\sigma ,\sigma '} \right)}
		\end{array}} \right)\left( {\begin{array}{*{20}{c}}
				{U\left( {E,\vec r\sigma '} \right)}\\
				{V\left( {E,\vec r\sigma '} \right)}
		\end{array}} \right)}  = \left( {\begin{array}{*{20}{c}}
			{E + \lambda }&0\\
			0&{E - \lambda }
	\end{array}} \right)\left( {\begin{array}{*{20}{c}}
			{U\left( {E,\vec r\sigma } \right)}\\
			{V\left( {E,\vec r\sigma } \right)}
	\end{array}} \right),
\end{equation}
where the particle-hole Hamiltonian $h$ and the one of pairing field $\tilde h$ have the local forms in the coordinate representation due to the zero-range interaction,
\beq
		h\left( {\vec r,\sigma ,\sigma '} \right) &=&  - \nabla M\nabla  + U + \frac{1}{{2i}}\sum\limits_{ij} {\left( {{\nabla _i}{\sigma _j}{B_{ij}} + {B_{ij}}{\nabla _i}{\sigma _j}} \right)} ,\nn
		\tilde h\left( {\vec r,\sigma ,\sigma '} \right) &=& {V_0}\left[ {1 - {V_1}{{\left( {\frac{\rho }{{{\rho _0}}}} \right)}^\gamma }} \right]\tilde \rho.
\eeq
Here, the terms of $M$, $U$ and $B_{ij}$ are related to the various local densities of nucleon and their explicit expressions are given in Ref.~\cite{stoisov05}. For a neutron drop system with axial symmetry, its density $\rho$ can be simply calculated by
\beq
	\rho \left( {r_{\perp},z} \right) = \sum\limits_k {{{\left| {{V_k}\left( {r_{\perp},z} \right)} \right|}^2}} ,
\eeq
where $r_{\perp}$ is the radial distance and $z$ is the height in cylindrical coordinate.

To investigate the shape of neutron drops, we calculate three kinds of root-mean-square (rms) radii, the total rms radii, $R_\text{tot}$, the axial rms radii, $R_{z}$, and the radial rms radii $R_{\perp}$, which are defined as:
\beq\label{zrrad}
		{R_{{\rm{tot}}}} &=&\sqrt{ \int {{{\vec r}^2}\rho \left( {r_{\perp},z} \right){d^3}\vec r}},\nn
		{R_z } &=&\sqrt{ \int {{z ^2}\rho \left( {r_{\perp} ,z} \right){d^3}\vec r}}, \nn
		R_{\perp} &=&\sqrt{ \int {{ r_{\perp}^2}\rho \left( {r_{\perp} ,z} \right){d^3}\vec r}}, 	
\eeq


It is difficult and time-consuming to solve above HFB equation in coordinate space. Therefore, the quasiparticle wave function is expanded in a complete set of basis functions which is generated by the eigenfunctions of an ASHO  potential:
\beq
		&{U_k}\left( {\vec r,\sigma } \right) = \sum\limits_\alpha  {{{\mathcal{U}}_{k\alpha }}{\Phi _\alpha }\left( {\vec r,\sigma } \right)} ,\nn
		&{V_k}\left( {\vec r,\sigma } \right) = \sum\limits_\alpha  {{\mathcal{V}_{k\alpha }}{\Phi _\alpha }\left( {\vec r,\sigma } \right)} ,
\eeq
where the wave functions and corresponding eigenvalues are explicitly written in the cylindrical coordinate as
\beq
	{\Phi _\alpha }\left( {\vec r,\sigma } \right) &=& \left[ {N_{{n_{\perp}}}^{{\Lambda}}{\beta _ {\perp} }\sqrt 2 {\eta ^{\left| {{\Lambda}} \right|/2}}{{\rm{e}}^{ - \eta /2}}L_{{n_{\perp}}}^{\left| {{\Lambda}} \right|}\left( \eta  \right)} \right]\left[ {{N_{{n_z}}}\beta _z^{1/2}{{\rm{e}}^{ - {\xi ^2}/2}}{H_{{n_z}}}\left( \xi  \right)} \right]\frac{{{{\rm{e}}^{i{\Lambda}\varphi }}}}{{\sqrt {2\pi } }}{\chi _{{\Sigma}}}\left( \sigma  \right),\nn
	{\varepsilon _\alpha } &=& \left( {2{n_{\perp} } + \left| {{\Lambda}} \right| + 1} \right)\hbar {\omega _{\perp} } + \left( {{n_z} + \frac{1}{2}} \right)\hbar {\omega _z}.
\eeq
$\Lambda$ and $\Sigma$ are the projections on the $z$-axis of the angular momentum and spin operators, respectively. The oscillator constants are defined as $\beta _\mu=\left(m\omega_\mu/\hbar\right)^{1/2}$ ($\mu=z,{\perp}$), and auxiliary variables, $\xi=z\beta_z$, $\eta=r_{\perp}^2\beta _{\perp} ^2$. ${{H_{{n_z}}}\left( \xi  \right)}$ and ${L_{{n_{\perp}}}^{ {{\Lambda}} }\left( \eta  \right)}$ represent the Hermite and associated Laguerre polynomials, respectively, with the normalization factors, ${N_{{n_{\perp}}}^{{\Lambda}}}$ and ${{N_{{n_z}}}}$. In practice, the size of basis is controlled through the parameter $N_{\text{sh}}={2{n_{\perp} } + \left| {{\Lambda}} \right| + n_z}$.

Solving the HFB equation (\ref{HFBeqn}) is then equivalent to diagonalize the matrix in each $\Omega^\pi$ block with $\Omega=\Lambda+\Sigma$ and $\pi=(-1)^{n_z+\Lambda}$,
\begin{equation}
	\left( {\begin{array}{*{20}{c}}
			{{\mathcal{H}} - \lambda }&\tilde{\mathcal{H}}\\
			\tilde{\mathcal{H}}&{ - {\mathcal{H}} + \lambda }
	\end{array}} \right)\left( {\begin{array}{*{20}{c}}
			{{\mathcal{U}_k}}\\
			{{\mathcal{V}_k}}
	\end{array}} \right) = {E_k}\left( {\begin{array}{*{20}{c}}
			{{\mathcal{U}_k}}\\
			{{\mathcal{V}_k}}
	\end{array}} \right),
\end{equation}
with the matrix elements
${\mathcal{H}}_{\alpha\beta}=\left\langle {{\Phi _\alpha }\left| {{h}} \right|{\Phi _\beta }} \right\rangle $ and $\tilde{\mathcal{H}}_{\alpha\beta}=\left\langle {{\Phi _\alpha }\left| {\tilde{h}} \right|{\Phi _\beta }} \right\rangle $.
}
 The more detailed formulas about Skyrme HFB model can be found in Ref.~\cite{stoisov05,dobaczewski84,bennaceur05}. The neutrons are trapped in an ASHO potential as,
\beq\label{ahop}
V_\text{ex}&=&\frac{1}{2}m\omega^2_{\perp}(x^2+y^2)+\frac{1}{2}m\omega^2_z z^2,\nn
&=&\frac{1}{2}m\omega^2_0r^2-m\omega^2_0r^2\delta\frac{4}{3}\sqrt{\frac{\pi}{5}} Y_{20},
\eeq
where $m$ is the neutron mass, with $\hbar^2/m=41.4$ MeV fm$^{2}$. The oscillator frequencies of the confinement potential in the radial and axial directions $\omega_{\perp}$ and $\omega_z$ measure the potential strengths in these directions respectively.  Alternatively, the ASHO can be expressed by an oscillator frequency of a spherical potential $\omega_0^2=(\omega^2_z+2\omega^2_{\perp})/3$ and a deformation parameter $\delta=3(\omega_{\perp}^2-\omega^2_z)/2(\omega^2_z+2\omega^2_{\perp})$ with a spherical harmonics $Y_{20}$.  When $\omega_{\perp}=\omega_z$, i. e., $\delta=0$, the external field is reduced to the spherical HO potential.  The deformation parameter $\delta$ measures the magnitude of the anisotropy of the ASHO potential.  The values $\delta=-1.5$ and $\delta=0.75$ correspond to the two extreme cases $\omega_{\perp}=0$ and $\omega_z=0$ respectively~\cite{ring80}.

\section{Numerical results and discussions}
In the calculation, the SLy4 parameter set is chosen as the particle-hole NN interaction and the mixed type of density-dependent delta interaction (DDDI) is taken for the particle-particle (pairing) channel as in Ref.~\cite{bennaceur05}.  The axially deformed neutron drop is calculated by the transformed HO basis program, HFBTHO~\cite{stoisov05}.  The spherical potential frequency $\hbar\omega_0$ is fixed as $10$ MeV.  In the following discussion, we focus on the neutron drop with $N_{\rm D}=28$, which is a traditional magic number in finite nuclei. Furthermore, its central density in a spherical HO potential with $\hbar\omega_0=10$ MeV is comparable to that of normal nuclei~\cite{ge2020a}.

\begin{figure}[htb]
	\centering
	\includegraphics[width=7cm]{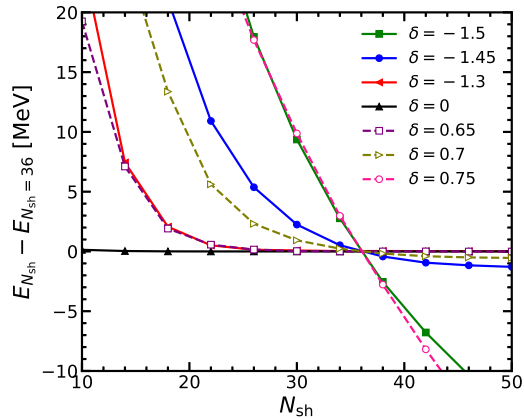}
	\caption{Convergence check of the total binding energy of neutron drop with the number of basis expansion shell, $N_{\rm sh}$, calculated in ASHO potential with different deformation parameters $\delta$, where the results calculated with $N_{\rm sh}=36$ are chosen as reference values.}
	\label{fig0}
\end{figure}

First, the convergence check of the total binding energy of neutron drop with the number of basis expansion shell, $N_{\rm sh}$ is carefully examined in the ASHO field with different deformation parameters $\delta$. The maximum $N_{\rm sh}$ in the program HFBTHO(v2.00d)~\cite{stoisov05} is $50$, which is large enough for normal nuclei. In Fig.~\ref{fig0}, the results of neutron drop calculated with $N_{\rm sh}=36$ are chosen as the reference values. It can be found that the binding energy in the spherical external field ($\delta=0$) converges very fast and becomes stable at $N_{\rm sh}=10$, which is similar to the case of normal nuclei. Such convergence is more difficult to achieve when the deformation of external ASHO potential becomes larger. 
At $\delta=-1.3$ and $\delta=0.65$, the results converge at around $N_{\rm sh}=30$, while for $\delta=-1.45$ and $\delta=0.7$ at almost the maximum $N_{\rm sh}=50$.
When the potential strength of one direction is zero, i.e. $\delta=-1.5$ or $\delta=0.75$, the calculated total binding energy of neutron drop decreases quickly with larger $N_{\rm sh}$, which cannot converge even with the maximum $N_{\rm sh}=50$. As a result, we cannot obtain the bound neutron drop for these extreme cases in the present calculation.  Actually, in these cases, the wavefunctions should be approximately regarded as plane waves in the radial or axial direction, which means that neutrons can move freely in these directions.  
A similar investigation of the ideal Fermi gas trapped in an anisotropic potential showed that the density of state of fermions $g(\epsilon)=\epsilon^2/(2\hbar^3\omega_z\omega^2_{\perp})$ will diverge in the case of $\omega_{\perp}=0$~\cite{giorgini08}.  With the maximum number of expansion shell $N_{\rm sh}=50$ in the present calculation, the lower and upper limits of the ASHO deformation parameters which can confine the neutron drop $N_{\rm D}=28$ are $\delta=-1.46$ and $\delta=0.72$, respectively.

\begin{figure*}[htb]
	\centering
	\includegraphics[width=15cm]{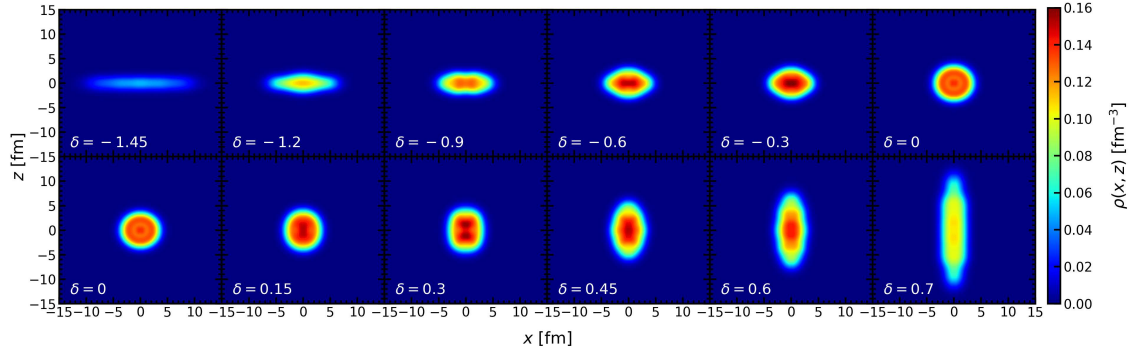}
	\caption{The shape evaluations of neutron drop, $N_{\rm D}=28$, in the $xz$ plane trapped in the ASHO field with $\delta=-1.45$ to $0.70$. In the upper panels, the neutron drops evaluate from oblate to spherical, while they will become prolate from spherical in the lower panels.}
	\label{fig1}
\end{figure*}
In Fig.~\ref{fig1}, the neutron density distributions in neutron drop, $N_{\rm D}=28$, trapped in ASHO potentials with $\delta=-1.45$ to $0.70$ are shown in the $xz$ plane.  It is clear that the shape of the neutron drop evolves from oblate to spherical and then to prolate as $\delta$ increases from negative to zero, and then to positive.  Specifically, at $\delta=-1.45$, the neutron drop is squeezed to be a disk-like structure as a quasi-two-dimensional system. Its density is very dilute, whose central value is just around $0.05$ fm$^{-3}$.  As $\delta$ increases, the neutron drop will be stretched in the axial direction.  It has a rod shape and approaches a quasi-one-dimensional system at $\delta=0.70$. It is interesting to mention that, for an open-shell neutron drop, it could be deformed collectively even in a spherical HO field with a smaller $\hbar\omega_0$,  when the pairing correlation is neglected as shown in Ref.~\cite{naito20}.   However, in the SHFB framework with pairing correlation, we have carefully examined that all the neutron drops $N_D=4$ to $N_D=50$ keep the spherical shapes at $\delta=0$.

\begin{figure*}[htb]
	\centering
	\includegraphics[width=15cm]{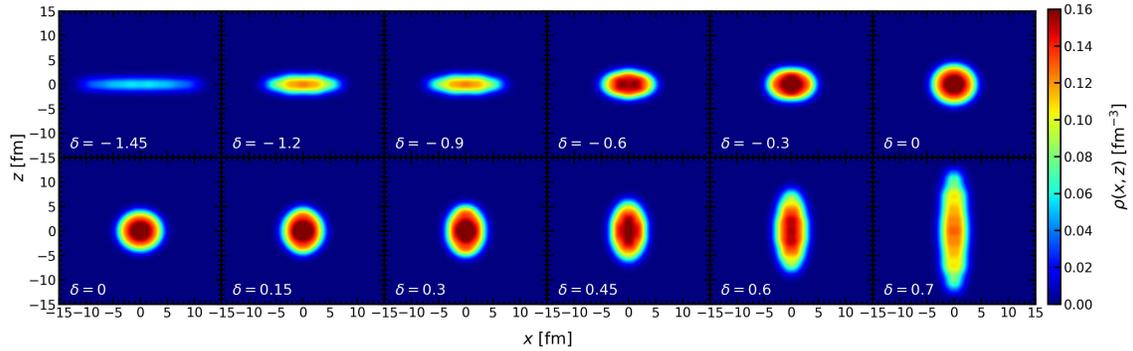}
	\caption{The shape evaluations of neutron drop, $N_{\rm D}=40$, in the $xz$ plane trapped in the ASHO field with $\delta=-1.45$ to $0.70$.}
	\label{fig1a}
\end{figure*}

Furthermore, the interior structure of the neutron drop is also largely influenced by the changes of $\delta$. In the spherical case $\delta=0$, the central density is slightly depressed just as in the `bubble' nuclei. As the anisotropy of the trap potential increases, some localization of neutron density happens at $\delta= 0.3$. The density of the two centers is much higher than that at the surface. 

{The case of neutron drop $N_\text{D}=40$ is also calculated and shown in Fig.~\ref{fig1a}. It can be found that the behaviors of a heavier neutron drop in the deformed external field are similar with those of $N_\text{D}=28$. The only difference is that the density in the central region becomes more compact with larger neutron numbers.}

\begin{table*}[htb]
	\centering
	\caption{The bulk properties of neutron drops, with $N_\text{D}=28$ and $N_\text{D}=40$, trapped in the ASHO potentials with different deformations, $\delta$.
		$E_\text{tot}$ presents the total energy, $R_\text{tot}$ the total root-mean-square (rms) radius,
		$R_{z}$ the rms radius along the symmetry axis, $R_{\perp}$ the rms radius in radial direction, $\beta$, the quadrupole deformation, and pairing gap $\Delta$. }\label{tab1}
	\resizebox{\textwidth}{!}{		
		\begin{tabular}{lcccccccccccc}
			\hline\hline
			\quad&$\delta $&~ $-1.45$ ~&~ $-1.20$ ~&~$-0.90$ ~&~$-0.60$ ~&~$-0.30$ ~&~$~0.00$  ~&~$~~0.15$ ~&~$~~0.30$  ~&~$~~0.45$ ~&~$~~0.60$  ~&~$~~0.70$  \\
			\hline
			\multirow{6}{*}{$N_\text{D}=28$}&  $E_\text{tot}/E_\text{HO}$&$0.83$&$0.75$&$0.73$&$0.72$&$0.72$&$0.71$&$0.72$&$0.72$&$0.73$&$0.73$&$0.75$\\
			&  $R_\text{tot}$ [fm]       &$7.12$&$4.58$&$3.76$&$3.40$&$3.28$&$3.17$&$3.23$&$3.31$&$3.54$&$4.16$&$5.73$\\
			&  $R_{z}$ [fm]              &$1.03$&$1.10$&$1.26$&$1.46$&$1.54$&$1.83$&$2.08$&$2.30$&$2.72$&$3.59$&$5.24$\\
			&  $R_{\perp}$ [fm]           &$6.85$&$4.45$&$3.55$&$3.08$&$2.89$&$2.59$&$2.47$&$2.38$&$2.27$&$2.10$&$1.95$\\
			&	$\beta$                   &$-0.74$&$-0.66$&$-0.53$&$-0.36$&$-0.27$&$ 0.00$&$ 0.19$&$ 0.35$&$ 0.61$&$ 0.98$&$ 1.31$\\
			&	$\Delta$ [MeV]            &$ 1.57$&$ 2.08$&$ 1.44$&$ 0.00$&$ 0.00$&$ 0.00$&$ 1.36$&$ 0.00$&$ 1.48$&$ 1.60$&$ 1.50$\\
			\hline	
			\multirow{6}{*}{$N_\text{D}=40$}&  $E_\text{tot}/E_\text{HO}$&$0.82$&$0.74$&$0.72$&$0.72$&$0.72$&$0.72$&$0.72$&$0.72$&$0.72$&$0.72$&$0.74$\\
			&  $R_\text{tot}$ [fm]       &$7.58$&$4.80$&$4.09$&$3.71$&$3.53$&$3.48$&$3.49$&$3.56$&$3.78$&$4.45$&$6.11$\\
			&  $R_{z}$ [fm]              &$1.02$&$1.16$&$1.29$&$1.49$&$1.72$&$2.01$&$2.22$&$2.50$&$2.91$&$3.85$&$5.77$\\
			&  $R_{\perp}$ [fm]           &$7.51$&$4.66$&$3.88$&$3.39$&$3.08$&$2.84$&$2.70$&$2.54$&$2.42$&$2.23$&$2.02$\\
			&	$\beta$                   &$-0.75$&$-0.65$&$-0.56$&$-0.41$&$-0.23$&$ 0.00$&$ 0.17$&$ 0.37$&$ 0.61$&$ 0.99$&$ 1.32$\\
			&	$\Delta$ [MeV]            &$1.63$&$1.77$&$1.38$&$1.47$&$1.72$&$0.92$&$1.63$&$1.14$&$1.52$&$1.28$&$1.60$\\
			\hline\hline
	\end{tabular}}
\end{table*}

In Table~\ref{tab1}, the bulk properties of neutron drop trapped in different ASHO potentials corresponding to Figs.~\ref{fig1} and \ref{fig1a}, such as its total energy,  $E_\text{tot}$, total root-mean-square (rms) radius, $R_\text{rms}$, $z-$axis rms radius, $R_z$, radial direction rms radius, $R_{\perp}$  quadrupole deformation, $\beta$, and average pairing gap $\Delta$ are listed. The total energy is scaled by the total energy of non-interacting nucleons trapped in the same external field, $E_\text{HO}=\sum^{N_{\rm D}}_{i=1}\varepsilon_i$ following the idea of spherical case.  Here the energy $\varepsilon_i$ of single-particle level in the ASHO potential in Eq.~(\ref{ahop}) has the analytical form~\cite{ring80},
\beq\label{ahoe}
\varepsilon_i=\hbar\omega_z\left(n_z+\frac{1}{2}\right) +\hbar\omega_{\perp}\left(N-n_z+1\right), ~~i=[N, n_z]
\eeq
where $N=2(n-1)+l$ is the principal quantum number of the oscillator major shell and $n_z$ is the quantum number along $z$-axis.
The ratios $E_\text{tot}/E_\text{HO}$ are around $0.71-0.83$, which means that the additional strong interactions make the neutrons more bound comparing to the non-interacting ones in the ASHO potential.  
This neutron drop $N_{\rm D}=28$ is most bound in the spherical external field with $\delta=0$. These values are similar to those of spherical neutron drops which are around $0.8$~\cite{zhao16,ge2020a,ge2020b}. The total rms radius of neutron drop becomes larger when the external field is more anisotropic because the potential will stretch or compress the neutron drops at the radial or axial direction. 
{The rms radii in the axial and radial directions with different $\delta$ are also calculated with Eq.~\ref{zrrad}. With the $\delta$ increasing, the rms radii in axial direction become larger from $1.03$ fm to $5.24$ fm, while the corresponding ones in the radial direction are largely reduced from $6.85$ fm to $1.95$ fm. This vividly shows that the neutron drop is stretched along the axial direction with positive increasing $\delta$. Furthermore, in the spherical external field, the $R_{\perp}=\sqrt{2}R_z$ is obtained self-consistently.}  

The quadrupole deformations of these neutron drops are from $-0.74$ to $1.31$, which are strongly dependent on the anisotropy of the external field.  It is easy to understand that the average pairing gap is zero at $\delta=0$ in the neutron drop $N_{\rm D}=28$ as a traditional magic number.  Besides, the zero average pairing gaps at $\delta=-0.60$, $-0.30$, and $0.30$ also indicate that the shell closure at $N_{\rm D}=28$ is prominent in these cases. Moreover, it is interesting to see that the average pairing gaps at other $\delta$ are not zero, which means that the shell structure is changed there due to the deformation.  

{The bulk properties of neutron drop, $N_{\rm D}=40$ are also shown in the lower panel of  Table ~\ref{tab1}. Its scaled energies, $E_\text{total}/E_\text{HO}$ and quadrupole deformations are very similar to those of $N_{\rm D}=28$. Its various rms radii become larger due to the increment of neutron numbers. Meanwhile, since the $N=40$ is not a traditional magic number, there is a finite pairing gap in the spherical external field due to the strong shell effect. }

\begin{figure}[htb]
	\centering
	\includegraphics[width=9cm]{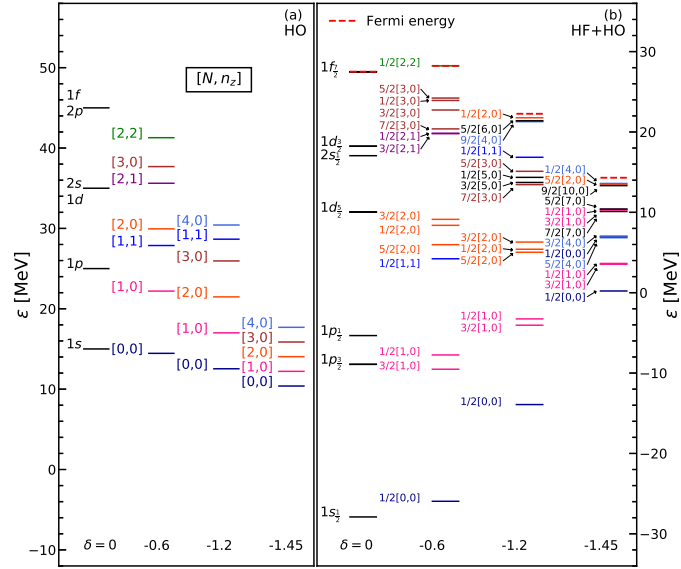}
	\caption{The energy levels of neutron systems, $N_{\rm D}=28$ confined in the ASHO potential with the deformed parameters $-1.45\leq\delta\leq0$. The energy levels of the non-interacting neutrons and those calculated with effective NN interactions SLy4 trapped in the same potentials are shown in panel (a) and panel (b) respectively.}
	\label{fig2}
\end{figure}

To better understand the shell evolution of the neutron drop $N_{\rm D}=28$ as a function of the anisotropy of the external field, we first present the levels of non-interacting neutrons in the ASHO potentials as given in Eq.~(\ref{ahoe}) at $\delta=0, -0.6, -1.2, -1.45$ in Fig.~\ref{fig2}(a).
At $\delta=0$, the $(1d,~2s)$ and $(1f,~2p)$ are degenerate in the major shells $N=2$ and $N=3$ respectively.  
As $\delta$ decreases, i. e., the potential strengths at the axial direction increases,  the levels with different $N$ and $n_z$ split separately.  Specifically, the levels with larger $N$ and smaller $n_z$ move down.  For example, the level of $[3,0]$ is lower than the one of $[2,2]$ at $\delta=-0.6$.  At $\delta=-1.2$, higher major shell states $[N,0]$ intrude down to the lower energy region.  More neutrons occupying the states with $n_z=0$ means they favor the state perpendicular to the symmetry axis, so-called `$\pi$ orbital'~\cite{itagaki01,itagaki04}.  More interestingly, all the neutrons occupy the energy levels $[N, 0]$ from $N=0$ to $N=4$ at $\delta=-1.45$, since $\hbar\omega_z$ is much larger than $\hbar\omega_{\perp}$, so the smallest $n_z=0$ could produce the lower energy, according to Eq.~(\ref{ahoe}). In this case, the energy levels have an equal interval, $\Delta E=\hbar\omega_{\perp}$.

The energy levels of neutron drop calculated with the Skyrme NN interaction are shown in panel (b) of Fig.~\ref{fig2}.  At $\delta=0$, the trap potential is spherical and thus the neutron drop is also spherical.  The spin-orbit partners like $(1p_{1/2},~1p_{3/2})$ and $(1d_{3/2},~1d_{5/2})$ have obvious energy splittings.  In normal nuclei with $N_{\rm A}=28$, the energy level of  $1s_{1/2}$ state is usually around $-50$ MeV. Because of the strong repulsive external field and the deficiency of protons, it increases to $-28$ MeV in neutron drops. Other energy levels also shift upward with similar magnitudes.  
Referring to Table~\ref{tab1}, the average pairing gap of this neutron drop at $\delta=0$ is zero.  Therefore, the full filling of $1f_{7/2}$ level makes the density increase more in the outer region, which leads to the relatively depressed density at the center and forms the `bubble' structure as shown in Fig.~\ref{fig1}.  

As $\delta$ decreases, the neutron drop becomes oblate as shown in Table~\ref{tab1}.   Conventionally, the energy levels of deformed nuclei are labeled by the asymptotic Nilsson quantum numbers $\Omega^\pi[N n_z \Lambda]$. 
It must be emphasized that the Nilsson quantum numbers obtained from the basis expansion method are those of the most important basis.  Here, we would like to apply the symbol $\Omega[N,n_z]$ ($\Omega$ is the projection of the total angular momentum) to denote the energy levels in deformed neutron drops, which is also convenient to compare with the energy levels of non-interacting neutrons in the same ASHO potential shown in Fig.~\ref{fig2} (a). 

When $\delta$ decreases from $0$ to $-1.45$, the lowest level of neutron drop moves upward.  This behavior is opposite to that of non-interacting neutrons in Fig.~\ref{fig2}(a). This is because the HF potential determined by the neutron density self-consistently becomes shallower when the neutroxn density is compressed farther away in the $xy$ plane as $\delta$ decreases from $0$ as shown in Fig.~\ref{fig1}.
Meanwhile, similar to the non-interacting neutrons in Fig.~\ref{fig2} (a), higher major shell states $[N,0]$ intrude down to the lower energy region as $\delta$ decreases. Therefore, the energy gaps among the levels with different $N$ obviously decrease, which makes the pairing scattering easier.  This can explain the increase of the average pairing gap from $0$ to $2$~MeV shown in Table~\ref{tab1}.

It is interesting to find that the energy levels of neutron drops calculated with the NN interactions are quite similar to those of non-interacting neutrons trapped in the same potentials, except that there are some splittings among the levels with the same $[N,n_z]$ in neutron drop due to the NN interaction at $\delta=-0.6$. 
These energy splittings are largely reduced at $\delta=-1.2$.  When the confinement of neutron drops at radial direction almost disappears, i.e., $\omega_{\perp}<<\omega_z$, for example, $\delta=-1.45$, the energy levels with the same $N$ have highly degenerated. The energy intervals of neighbor main shells are almost the same, which is similar to the case of non-interacting neutrons shown in (a) panel. 

At the other extreme case, when the neutron drop is stretched to a rod shape, i.e., $\omega_z<<\omega_{\perp}$, the total energy will be smaller with $n_z=N$ according to Eq.~(\ref{ahoe}).  This means that the neutrons prefer to occupy the orbital parallel to the symmetry axis, the so-called `$\sigma$ orbital'.  This is consistent with the available investigations in the chain structure of carbon isotopes by Itagaki {\it{et al.}} and Zhao {\it {et al.}}~\cite{itagaki01,itagaki04,zhao15}. 

\section{Summary}
In summary, the neutron drop trapped in an ASHO potential was firstly investigated within the Skyrme HFB model.  The properties of neutron drop, such as the total energy, deformation, rms radius, and density distribution are strongly dependent on the anisotropy of the ASHO potential.  When the ASHO potential strength at the axial direction is much larger than the one in the radial direction, the neutron drop is suppressed as a disk-like shape. However, once the ASHO potential strength at the axial or radial direction completely disappears, the neutrons will scatter out like the non-interacting neutron case.

Generally speaking, the energy levels of neutron drop are highly analogous to those of the non-interacting neutrons in the ASHO potential.  The addition of NN interaction leads to the splittings among levels with same quantum numbers $[N,n_z]$. The levels with large $N$ and small $n_z$ shift down quickly as the potential strength of the axial direction increases, which means that neutrons favor the levels perpendicular to the symmetry axis.
Besides, there is a rich interior structure in neutron drops, such as 'bubble' or 'localization' which is also dependent on the anisotropy of the external potential. In the rod-shaped neutron drop, the neutrons prefer to occupy the $\sigma$ orbital. These investigations indicate that the manifold phenomenons in a nuclear many-body system can be realized in a neutron drop by controlling the anisotropy of an external field, just like the ultracold atom.

\section*{Acknowledge}
The author J. N. Hu and Y. Zhang would like to thank all the friends' warm greetings and encouragements when they were trapped in hotels in Shanghai after they came back to China from RIKEN.  They are also grateful for this isolated time to focus on this paper about the trapped neutron drop. This work was supported in part by the National Natural Science Foundation of China (Grants  No. 11775119, No. 11675083, and No. 11405116),  the Natural Science Foundation of Tianjin, and China Scholarship Council (Grant No. 201906205013 and No. 201906255002).

\end{document}